\documentclass[12pt]{iopart}

\usepackage{graphicx}
\usepackage[utf8x]{inputenc}

\begin{document}


\title{Vortex dynamics and transport phenomena in stenotic aortic models using Echo-PIV}

\author{Javier Brum$^1$, Miguel Bernal$^2$\footnote{Present address:  Verasonics Inc,11335 NE 122nd Way, Suite 100, Kirkland, WA 98034, USA},
Nicasio Barrere$^3$, Carlos Negreira$^1$ and Cecilia Cabeza$^3$}

\address{$^1$ Laboratorio de Acústica Ultrasonora, Instituto de Física, Facultad de Ciencias, Universidad de la República, Iguá 4225, 11400, Montevideo, Uruguay}
\address{$^2$ Grupo de Dinámica Cardiovascular, Universidad Pontificia Bolivariana, Medellín, Colombia}
\address{$^3$ Física No Lineal, Instituto de Física, Facultad de Ciencias, Universidad de la República, Iguá 4225, 11400, Montevideo, Uruguay}
\ead{jbrum@fisica.edu.uy}


\date{\today}

\begin{abstract}

Atherosclerosis is the most fatal cardiovascular disease. As disease progresses, stenoses grow inside the arteries blocking their lumen and altering blood flow. Analysing flow dynamics can provide a deeper insight on the stenosis evolution. In this work, we propose a novel approach which combines ultrasound with Eulerian and Lagrangian descriptors, to analyse blood flow dynamics and fluid transport in stenotic aortic models with morphology, mechanical and optical properties close to those of real arteries. To this end, vorticity, particle residence time (PRT), particle's final position (FP) and finite time Lyapunov's exponents (FTLE) were computed from the experimental fluid velocity fields acquired using ultrasonic particle imaging velocimetry (Echo-PIV). For the experiments, CT-images were used to create morphological realistic models of the descending aorta with 0\%, 35\% and 50\% occlusion degree with same mechanical properties as real arteries. Each model was connected to a circuit with a pulsatile programmable pump which mimics physiological flow and pressure conditions. The pulsatile frequency was set to $\approx$ 0.9 Hz (55 bpm) and the upstream peak Reynolds number ($Re$) was changed from 1100 to 2000. Flow in the post-stenotic region was composed of two main structures: a high velocity jet over the stenosis throat and a recirculation region behind the stenosis where vortex form and shed. We characterized vortex kinematics showing that vortex propagation velocity increases with $Re$. Moreover, from the FTLE field we identified Lagrangian Coherent Structures (i.e. material barriers) that dictate transport behind the stenosis. The size and strength of those barriers increased with $Re$ and the occlusion degree. Finally, from the PRT and FP, we showed that independently of $Re$, the same amount of fluid remains on the stenosis over more than a pulsatile period, which combined with large FTLE values may provide an alternative way to understand stenosis growth. 

\end{abstract}


\submitto{\PMB}

\maketitle


\section{Introduction}

Cardiovascular diseases represent one of the major causes of death in the world \cite{mendis2011global}. Blood flow dynamics has shown to be crucial in understanding the different causes of gradual or acute changes during diseases affecting the cardiovascular system. For example, flow pattern and vortex dynamics in the left ventricle are correlated with altered cardiac function \cite{bermejo2014intraventricular, hong2008VortexCharacterization,faludi2010left}. In arteries, regions of disrupted flow or with large recirculation are likely to favor atherosclerosis \cite{zarins1983carotid,martorell2014extent}, one of
the most prevalent and dangerous cardiovascular disease \cite{mendis2011global}. In atherosclerosis fat, cholesterol and other substances accumulate inside the artery creating a plaque or stenosis which narrows the arterial lumen. During growth or plaque rupture, atherosclerosis may lead to heart failure, stroke or even death. However, despite its danger, the mechanisms of stenosis evolution and rupture remain poorly understood. 

Eulerian approaches have shown to be a powerful tool in analysing blood flow dynamics in stenotic vessels \cite{varghese2007direct, katritsis2010vortex,choi2018flow, geoghegan2013time, pielhop2012analysis,usmani2016pulsatile}. In these works, streamlines, vorticity, turbulent kinetic energy and wall shear stresses were computed from the instantaneous velocity data obtained either via numerical studies \cite{varghese2007direct, katritsis2010vortex}, in vitro experiments \cite{choi2018flow, geoghegan2013time, pielhop2012analysis} or both \cite{usmani2016pulsatile}. Alternatively, Lagrangian descriptors can provide direct information on the transport topology in large vessels (e.g. flow mixing, stirring, recirculation, stagnation and separation) \cite{shadden2008LCS_Vessels}. Analysing the transport topology in stenotic vessels may provide valuable information on how stenosis grows and eventually ruptures \cite{shadden2013potential, xu2009clotgrowth,rayz2010flow}. 

Lagrangian Coherent Structures (LCS) are the most common Lagrangian descriptors and are usually defined as locally strongest attracting or repelling material barrier \cite{haller2015lagrangian, shadden2012FTLEbook}. They reveal structures that govern fluid transport and mixing in complex flows and have been applied to oceanography, geophysics and atmospheric flows (please refer to references within the reviews \cite{haller2015lagrangian, shadden2012FTLEbook}). Another Lagrangian descriptor is the fluid Particle Residence Time (PRT), which measures the time that a fluid parcel spends in a given region. Recently LCS and PRT have been applied to study hemodynamics in the left ventricle \cite{espa2012lagrangian, rossini2016LVstasis, badas2017quantificationFTLE_pdf, diLabio2018material, hendabadi2013topology, toger2012vortex, charonko2013vortices} and large vessels \cite{babiker2012vitro, patel2017effect, jeronimo2019prt,jeronimo2020stenosis, rayz2010flow, vetel2009lagrangian, shadden2008LCS_Vessels, arzani2012FTLE_AAA}. 

However, despite the numerous works involving hemodynamics in pathological conditions, only a few have focused on studying flow dynamics in stenotic vessels. In the experimental work of Jerónimo \textit{et al.} \cite{jeronimo2019prt,jeronimo2020stenosis} PRT was measured using two dimensional particle tracking velocimetry in steady and unsteady flow conditions for different Reynolds and Strouhal numbers. As vessel model, they used a rigid acrylic pipe with a smooth axisymmetric constriction followed by an unrealistic sudden expansion. Other works used instantaneous velocity data obtained either via numerical studies \cite{varghese2007direct, katritsis2010vortex,usmani2016pulsatile} or in vitro experiments \cite{choi2018flow, geoghegan2013time, pielhop2012analysis, usmani2016pulsatile}. Numerical studies can account for patient specific morphology \cite{katritsis2010vortex}, providing highly resolved three-dimensional velocity fields. However, they are unable to simulate the mechanical properties of the vessel wall, which is always assumed to be rigid. This is also the case for the experiments \cite{choi2018flow, geoghegan2013time, pielhop2012analysis,usmani2016pulsatile} where transparent models made of acrylic or silicone are required because optical particle velocimetry is used to measure the fluid velocity field. For example, in the work of Usmani \textit{et al.} \cite{usmani2016pulsatile} they used a compliant stenotic model made of silicone where its distensibility was changed by modifying the wall thickness of the model. This type of material does not mimic the anisotropic, non-linear, elastic behaviour of the vessel wall. Finally, these works focus on studying flow from an Eulerian perspective. To our knowledge, no experimental work reported the use of LCS to study flow in stenotic vessels. 

Consequently, in this work we present a novel approach to study vortex dynamics and fluid transport in stenotic aortic models with morphology, mechanical and optical properties close to those of real arteries. To this end, vorticity, LCS, PRT and final position (FP) maps were computed from the particle velocity fields acquired using ultrasonic particle imaging velocimetry (Echo-PIV). The flow dynamics behind the stenosis was studied as a function of Reynolds number and the degree of occlusion. Finally, results were discussed and compared to those reported in the literature.

\section{Materials and Methods}

\subsection{Model Manufacturing}

The aortic models for this study were developed following the same procedure described in Bernal \textit{et al.}  \cite{bernal2019models}. By segmentation of anonymous CT-images of a healthy patient, the lumen and the adventitia layer of the descending aorta were reconstructed (lumen volume and inflated volume in Fig. 1(a)). For each geometry (lumen and inflated), two molds were fabricated using computer numerically controlled machine. Then, the lumen volume was inserted into the mold created from the inflated geometry (Fig. 1(b)). The gap between these two elements allowed the injection of a 10\% b/w Polyvinyl Alcohol (PVA) solution. Prior to the injection of the polymer, the core was wrapped with a reinforcing fabric to allow the models to withstand physiological pressure levels and to have mechanical properties similar to that of healthy arteries \cite{bernal2019anastomoses,bernal2019models}. Finally, the whole setting was subject to 7 cycles of freezing and thawing to polymerize PVA, after which the mold was opened and the core was removed. The models dimensions were 17 cm in length, 2.4 cm of external diameter and $\approx$ 0.3 cm of wall thickness. For healthy physiological pressures (diastolic/systolic) of 80/120 mmHg, the models exhibit a nonlinear change in their shear modulus from $\approx$ 133 kPa to $\approx$ 209 kPa with 10\% of error. These values are very similar to those of swine aortas \cite{bernal2019anastomoses,bernal2019models}. Precise details on the mechanical evaluation of the models are given in the works of Bernal \textit{et al.}  \cite{bernal2019anastomoses,bernal2019models}.

Additional models with different degree of occlusion were made by modifying the lumen geometry resulting in a stenosis made of PVA (Fig. 1(b)). The profile of the stenosis was chosen to be Gaussian following $z_{st} = A_oexp[-x/(2\sigma^2)]$ where $z_{st}$ is the height of the stenosis and $\sigma$ = 1.85 cm, which gives a full width at half maximum of $\approx$ 4.35 cm. The value of $A_o$ was chosen to achieve an occlusion maximum height of $\approx$ 0.8 cm and $\approx$ 1.2 cm which resulted in two different models with 35\% and 50\% of occlusion in diameter, respectively. The shear modulus of the stenosis was 30 kPa and was evaluated by shear wave elastography. Figures 1(c) and 1(d) show a cross sectional B-mode image of each of the stenotic models with the Gaussian profile highlighted by a full red line.

\begin{figure*}[!]
\includegraphics[width = 1 \linewidth]{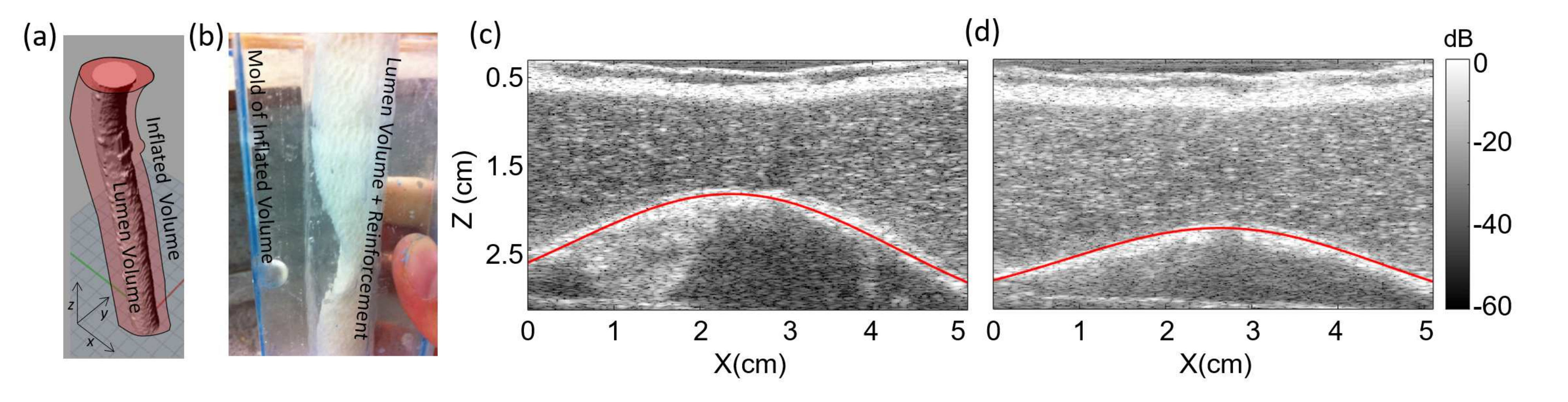}
\caption{\label{fig:models} Manufacturing process of the stenotic aortic models. \textbf{(a)} Reconstructed volumes from CT-images: lumen volume in grey and inflated volume in red shadow. \textbf{(b)} Lumen volume wrapped in the reinforcing fabric inside the acrylic mold created from the inflated volume. Cross-sectional B-mode images of the models presenting a \textbf{(c)} 50\% and \textbf{(d)} 35\% occlusion in diameter. The red full line corresponds to the Gaussian profile used to create the stenosis.}
\end{figure*}

\subsection{Hemodynamic simulator}

Each model was connected to a hemodynamic work bench simulator, which has already been used for the elasticity assessment under physiological conditions of vascular grafts, vessel models and femoral ovine arteries \cite{bagnasco2014elasticity, bernal2019models}. The simulator consists of a programmable piston pump that mimics flow and pressure wave from the heart \cite{balay2010improvement}, two pressure sensors (one at the inlet of the model and one at the outlet), a reservoir that could be pressurized using a manual sphygmomanometer and a water bath where the model is placed.  

\subsection{Echo-PIV}

The fluid velocity field was measured through ultrasonic particle imaging velocimetry (Echo-PIV), which is based on 2D cross-correlation of consecutive speckle images \cite{kim2004development, zheng2006real, jensen2016ultrasound_I}. To this end, the circuit was filled with degassed water seeded with neutrally buoyant polyamide particles (Dantec Dynamics, Denmark) with a mean diameter of 50 $\mu$m. These particles have a Stokes number of approximately $1 \times 10^{-4}$ and can be assumed to closely follow the flow. A particles density of 0.2 gr/L was used in all the experiments. 

For Echo-PIV, ultrasonic B-mode images were acquired using plane wave insonification \cite{montaldo2009coherent} at 200 Hz frame rate during 2.5 s. A custom-made linear probe (Vermon, Tours, France) with 256 elements (0.2 mm pitch) working at 15 MHz driven by a Verasonics Research Ultrasound System was used in the experiments. The axial resolution (i.e. along depth) of the B-modes images was 0.1 mm. The cross sectional centre plane of the model was imaged by positioning the probe parallel to the aortic model with its first element facing towards the direction of the flow (Fig. 2(a)). To avoid constraint the movement of the wall, the ultrasound probe was not in direct contact with the model using the water bath as coupling medium between probe and model. As illustrated in Fig. 2(a) by different coloured rectangles, three Regions of Interest (ROI) were imaged in the experiments: upstream (black dashed-dotted line), post-stenotic (red full line) and on the stenosis throat (blue dashed line). To image these three ROIs the probe was moved along the x-direction by a step by step motor. For each experiment, a time lapse of ten periods was waited before the ultrasound acquisition to avoid start-up effects. 

Finally, PIVlab software was used to compute the velocity fields from the B-mode images using a direct correlation approach with windows of 0.64 x 0.32 cm$^2$ and 75\% overlap \cite{thielicke2014pivlab}. This resulted in a lateral and axial resolution for the velocity fields of 1.6 mm and 0.8 mm, respectively.

\begin{figure*}[!]
\includegraphics[width = 1 \linewidth]{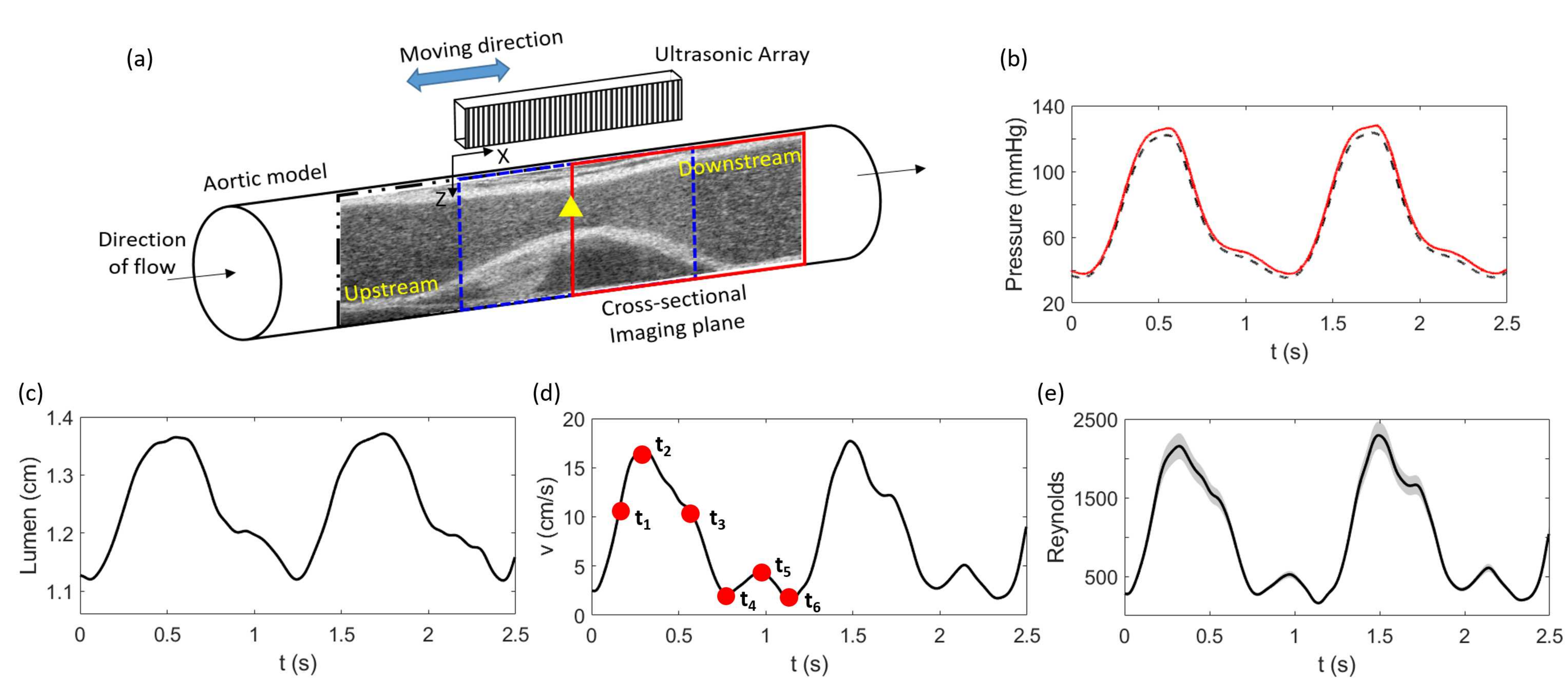}
\caption{\label{fig:setup} Echo-PIV setup and flow conditions. \textbf{(a)} Sketch of the ultrasonic probe positioning and movement along the cross-sectional centre plane of the aortic model (xz-plane). The three ROIs upstream, on the stenosis and post-stenotic are indicated by a black dotted-dashed, blue dashed and solid red rectangles, respectively. \textbf{(b)} Pressure at the inlet (red solid line) and outlet (black dashed line) of the 50\% occluded model for $Re$ = 2000 experiment. \textbf{(c)} Lumen size and \textbf{(d)} velocity as a function of time measured at the centreline of the stenosis throat indicated by a yellow triangle on panel (a). \textbf{(e)} $Re$ number computed from panels (c) and (d). The gray shadow indicates the error on $Re$ due to the error on the diameter which is $\approx$ 0.1 cm.}
\end{figure*}

\subsection{Flow Conditions}

The pulsatile frequency was set to $\approx$ 0.9 Hz (55 bpm) for all the experiments, close to that of a normal human heart. Moreover, the pressure wave form was set to be close to the phsysiological pressure wave for the descending aorta \cite{o1968pressure}. Figure 2(b) shows representative pressures at the inlet and outlet of the 50\% occluded model.  

Given a fixed flow rate, flow dynamics is essentially controlled by the Reynolds number ($Re$), which relates the flow velocity ($v$) with lumen size ($D$) as $Re = Dv/\nu$, where $\nu$ is the kinematic viscosity of water ($1.0 \times 10^{-6}$ m$^2$/s). Taking into account the internal diameter of our models and the kinematic viscosity of water, the pulsatile pump was programmed to achieve three different peak velocities inside the models: 9.5 cm/s, 7 cm/s and 5 cm/s. This corresponds to peak Reynolds numbers of approximately 2000, 1500 and 1100, respectively, which lie within the physiological range of the aortic artery \cite{ha2018age_aorta}. Nevertheless, small deviations from these values are expected in the experiments because of the small differences in  the mechanical response of the models. Therefore, the exact Reynolds number for each experiment was computed by measuring the internal diameter and the velocity obtained through Echo-PIV. As an example, Figs. 2(c) to 2(e) show the lumen, velocity and $Re$ as a function of time for the $Re$ = 2000 experiment in the 50\% occluded model. Velocity and diameter are affected by a 5 \% error which results in an uncertainty of 10\% for $Re$ (grey shadow in Fig. 2(e)). The internal diameter of the model was estimated by tracking the movement of the wall from the B-mode images using an intensity threshold of -6 dB. A summary of the precise experimental conditions is given in Table 1. Repeatability of the experiments, models' mechanical properties and flow patterns was checked from experiments carried out over a five-month period. 

In Table 1, the values of "$Re$" set on the programmable pump and  $Re_{max}$ agree within the margins of error for all the experiments. Therefore, in this work we will use the set value to label each experiment. Finally, it is important to mention that the pressure wave-form and the shape of the velocity traces presented in Fig. 1 were preserved for all experiments defining a time reference indicated by the numbered red circles in Fig. 2(d). Times $t_1$ to $t_6$ correspond to 0.25$T$, 0.40$T$, 0.55$T$, 0.70$T$, 0.85$T$ and $T$,respectively, being $T$ the pulsatile period.

\begin{table}
\caption{\label{expcond}Summary of the experimental flow conditions.}
\footnotesize
\centering
\resizebox{\textwidth}{!}{\begin{tabular}{@{}ccccccccccc}
\br
occlusion & $Re$ $^\dagger$   &        & $v_{max}$ (cm/s) & $v_{min}$ (cm/s) & $D_{max}$ (cm) & $D_{min}$ (cm) & $Re_{max}$ & $Re_{min}$ & $P_{max}$ (mmHg) $^\diamond$ & $P_{min}$ (mmHg) $^\diamond$\\
\mr
50\%      & 2000 & inlet  & 8.11             & 0.29             & 2.49           & 2.37           & 1950       & 100        & 128            & 37               \\
          &      & throat & 17.7             & 1.39             & 1.37           & 1.12           & 2290       & 160        & 123            & 35               \\
          & 1500 & inlet  & 6.13             & 0.20             & 2.37           & 2.28           & 1422       & 47         & 64             & 24               \\
          &      & throat & 12.1             & 0.49             & 1.22           & 1.06           & 1401       & 55         & 59             & 22               \\
          & 1100 & inlet  & 4.77             & 0.20             & 2.31           & 2.19           & 1087       & 45         & 43             & 18               \\
          &      & throat & 9.73             & 0.39             & 1.13           & 0.98           & 1034       & 43         & 40             & 17               \\
35\%      & 2000 & inlet  & 8.39             & 0.21             & 2.54           & 24.03          & 1983       & 50         & 95             & 18               \\
          &      & throat & 12.67            & 0.51             & 1.77           & 1.50           & 1963       & 81         & 92             & 18               \\
          & 1500 & inlet  & 7.19             & 0.32             & 2.43           & 2.24           & 1694       & 71         & 45             & 11               \\
          &      & throat & 9.12             & 0.28             & 1.64           & 1.39           & 1424       & 43         & 43             & 11               \\
          & 1100 & inlet  & 5.74             & 0.28             & 2.35           & 2.18           & 1288       & 65         & 27             & 8                \\
          &      & throat & 7.58             & 0.22             & 1.55           & 1.31           & 1107       & 32          & 25             & 8                \\
0\%       & 2000 & inlet   & 7.64             & 0.49             & 2.41           & 2.29           & 1806       & 112        & 133            & 32                

\end{tabular}}\\
$^\dagger$ This value corresponds to the $Re$ number set on the programmable pump. $^\diamond$ For $P_{max}$ and $P_{min}$, the rows "inlet" and "throat" correspond to inlet and outlet of the model respectively.

\end{table}
\normalsize



\subsection{LCS identification} 

A common method to identify Lagrangian Coherent Structures (LCS) is by computing the Finite-Time Lyapunov Exponents (FTLE). As demonstrated in \cite{shadden2005definition} the surfaces that maximize the FTLE field (i.e. ridges) correspond to the LCSs. FTLE measure the rate of separation between initially adjacent particles that are advected by the flow over a finite time interval ($\tau$). Therefore, LCS will represent surfaces of large particle separation, which act as a material barrier identifying regions with different flow dynamics. 

To compute the FTLE field the first step consists in seeding the fluid domain with a grid of particles. These particles will be advected from $t_0$ to $t_0 + \tau$ by integrating the particle velocity field over time. This gives the flow map $\phi^{t_0+\tau}_{t_0}: r(t_0) \rightarrow r(t_0 + \tau)$, where $r$ denotes the particle position. The amount of stretching about a trajectory can be defined in terms of the Cauchy-Green tensor

\begin{equation}
C(r_0,t_0,\tau) = \bigtriangledown \phi^{t_0+\tau}_{t_0}(r_0)^\dagger .\bigtriangledown \phi^{t_0+\tau}_{t_0}(r_0)
\end{equation}
evaluated at the initial position $r_0 = r(t_0)$. The maximum stretching $\left \| \delta r \right \|_{max} (= \exp^{\left |  \tau \right | \Lambda}\left \| \delta r_0 \right \|)$ is aligned with the eigenvector of maximum eigenvalue $\lambda_{max}$ of $C$. Consequently, the FTLE $\Lambda$ may be computed as:

\begin{equation}
\Lambda (r_0,t_0,\tau) = \left |  \tau^{-1} \right | \ln{\sqrt{\lambda_{max}((r_0,t_0,\tau))}}
\end{equation}
For unsteady flows, Eq. 2 is computed for a range of times $t_0$ to provide a time series of the FTLE allowing to follow the dynamics of the LCSs.

Given $\tau$, trajectories can be integrated forward ($\tau > 0$) or backward ($\tau < 0$) in time. For a positive $\tau$, the FTLEs measure the rate of separation, thus identifying repelling structures ($\Lambda^+$ field). On the contrary, if $\tau$ is negative, the FTLEs measure the rate of convergence, thus identifying attracting structures ($\Lambda^-$ field). The choice of $\tau$ is usually related to the characteristic flow time scales. In this work a characteristic time scale is imposed by the period ($T$) of the pulsatile flow. Another characteristic time scale is given by the advective time in the post-stenotic region ($D/v \sim 0.3-0.5$ s). In light of both time scales, a $\tau = T/3 = 0.35$ s was used in this work. 

To obtain sharper fields, FTLEs are usually computed in a grid finer than the original grid used for the particle velocity field. In this work the grid of the particle velocity data was subdivided to contain $3 \times 3$ fluid particles. Particles were advected using a 4th order Runge-Kutta method with a cubic interpolation. Further refinement of the particle grid resulted in negligible change in the results. Lastly, as LCSs (i.e. ridges of the FTLE field) we considered all the FTLEs whose value exceeded a threshold of 50\% of the maximum value of the FTLE field (this threshold value is larger than the field mean value plus three times its standard deviation).  

\subsection{Residence time and final position mapping} 

Particle Residence Time (PRT) maps have shown to be a powerful lagrangian tool when analysing transport phenomena. Each pixel of the map corresponds to the fluid particle's initial position and the pixel value is given by the time spent by this particle within the ROI. To complete the information given by the PRT maps, in this work we introduced the Final Position (FP) maps. Analogous to PRT maps, each pixel of the FP map corresponds to the particle's initial position. However, for the FP maps the ROI is subdivided in different color-coded subregions. Then, the pixel value is determined by the fluid particle's final position within these subregions. In this work the post-stenotic ROI was subdivided into three different subregions (please refer to section \ref{prt_results} for the definition of subregions). For the PRT and FP maps the beginning of the systolic phase ($t = 0$ in Fig. 2(d)) was considered as the initial time. Fluid particles were advected over one pulsatile period following the same procedure described in the preceding subsection.

\begin{figure*}[t!]
\includegraphics[width = 1 \linewidth]{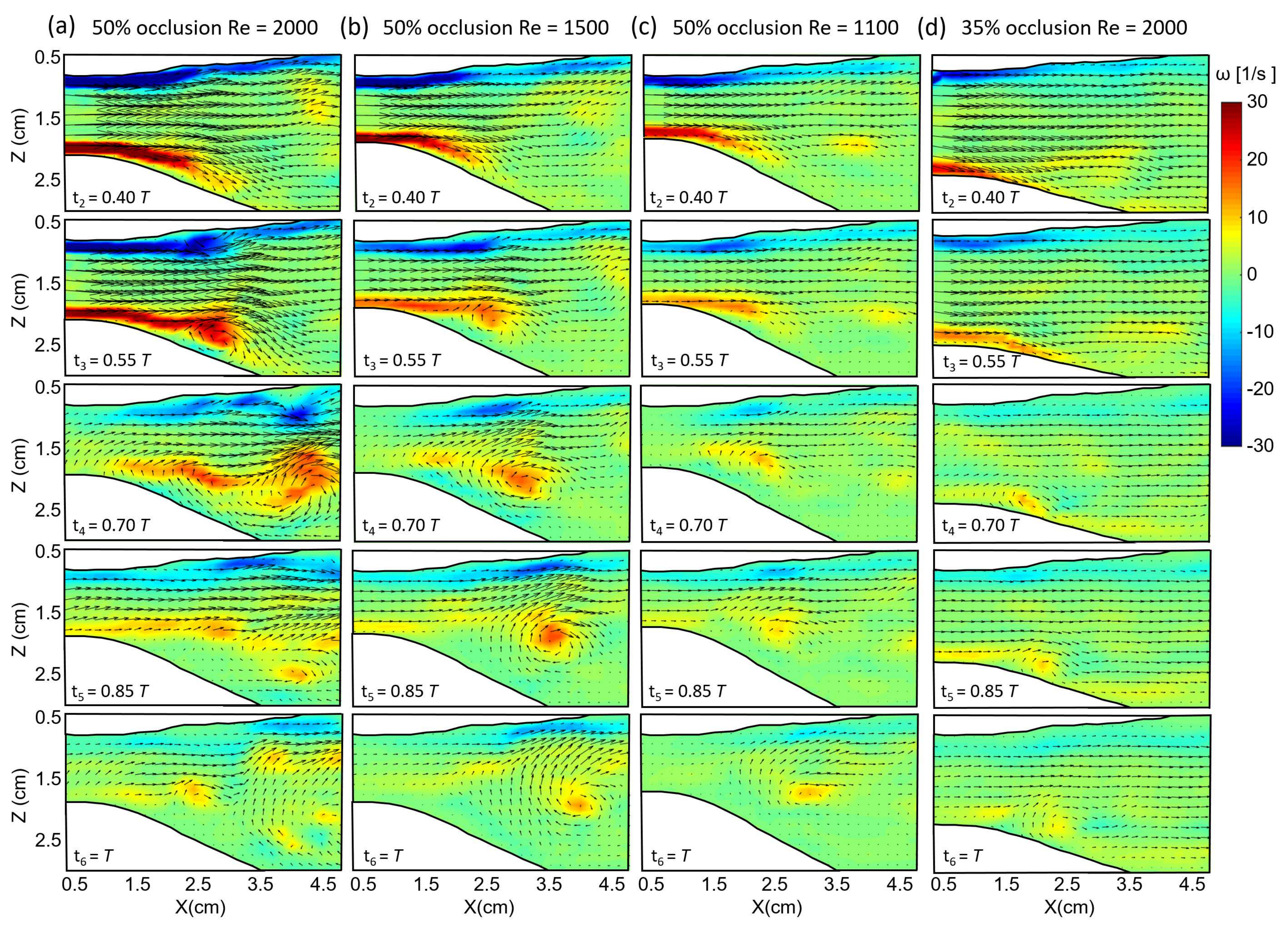}
\caption{\label{fig:vorticity} Five snapshots of the instantaneous particle velocity field superimposed over the vorticity field $\omega$ for the following experiments \textbf{(a)} 50\% occlusion, $Re$ = 2000, \textbf{(b)} 50\% occlusion, $Re$ = 1500, \textbf{(c)} 50\% occlusion, $Re$ = 1100 and \textbf{(d)} 35\% occlusion, $Re$ = 2000. Each snapshot corresponds to the times $t_2$ to $t_6$ indicated in Fig. 2(d) by the red dots. The color bar is the same for all panels.}
\end{figure*}

\section{Results}

\subsection{Eulerian description of the flow: particle velocity and vorticity fields}

Flow in the unobstructed model and in the upstream region of the occluded models showed a laminar velocity profile without any flow instabilities (not shown). However, in the post-stenotic region of the occluded models, flow transitioned from laminar to vortex formation and shedding as $Re$ increased. Figure 3 shows five snapshots (times $t_2$ to $t_6$ in Fig. 2(d)) of the instantaneous velocity field superimposed over the vorticity $\omega = \nabla \times v$ for different experiments. The main features of the flow are highlighted by this figure: a high velocity jet over the stenosis throat and large recirculation region behind the stenosis where vortex form and shed. For a 35\% occlusion, at $Re$ = 2000 a small recirculation region without shedding appeared behind the stenosis (Fig. 3(d)). Below this $Re$ number flow was mostly laminar. Contrary, for the model with a 50\% occlusion, vortex shedding happens for all the $Re$ (Fig. 3 (a)-(c)). As expected, as $Re$ increased the vortices become larger and vorticity values higher. By comparing Figs. 3(a)-(c) vortices not only become larger and stronger but they also propagate faster as $Re$ increased. By tracking the centre of the vortex as a function of time, vortex propagation velocities
of 1.9 cm/s, 2.8 cm/s and 6.2 cm/s were found for $Re$ = 1100, 1500 and 2000 respectively.

\begin{figure*}[t!]
\includegraphics[width = 1 \linewidth]{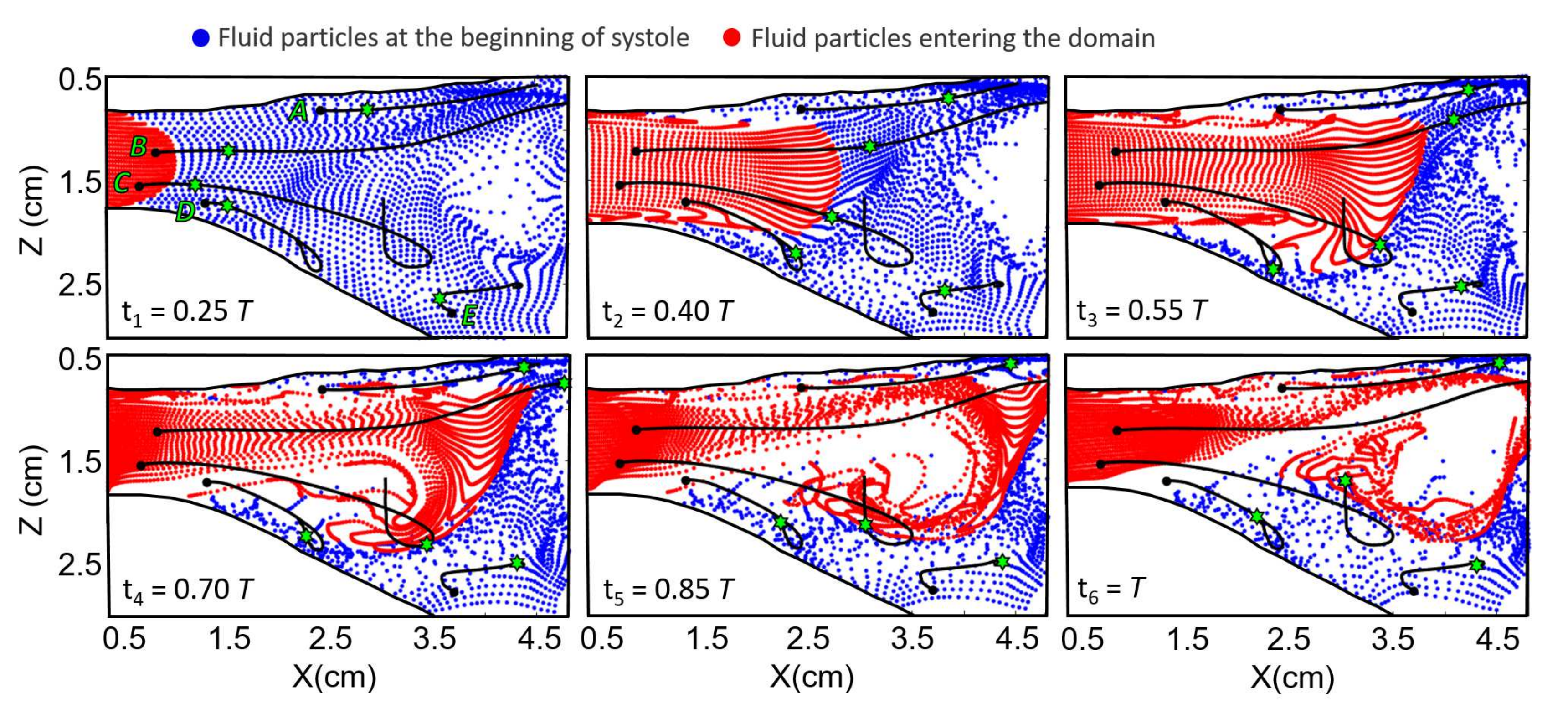}
\caption{\label{fig:partilces} Six snapshots of the particle advection in the 50\% occluded model for $Re = 1500$. Particles present at the beginning of the systole ($t=0$ in Fig 2(d)) are represented in blue dots, while particles entering the domain during one pulsatile period are represented in red. The time reference corresponding to each snapshot is indicated on the left lower corner of each panel. Additionally, in each panel different particle's trajectories (labeled $A$ to $E$) are represented by black full lines. The full black circle indicates the particle's initial position while the green star represents their position in each snapshot.}
\end{figure*}

\subsection{Lagrangian description of the flow}

As shown in Fig. 3 the main structures that govern flow behind the stenosis are the high velocity jet and the vortex. Consequently, those structures will be also responsible for dictating the transport phenomena in the post-stenotic region. Figure 4 shows six snapshots of the fluid particle's advection in the 50\% occluded model for $Re = 1500$. Fluid particles at the beginning of the systole (i.e. pre-existing fluid) are represented in blue dots, while fluid particles entering the domain during one pulsatile period are represented in red.

The main features of the fluid transport are highlighted by Fig. 4. As flow enters the domain through the jet, those fluid particles situated at the leading edge of the jet do not mix with the pre-existing fluid. Contrary, mixing occurs mainly in the recirculation zone behind the stenosis where pre-existing fluid is dragged and carried away by the vortex. Finally, there are fluid particles that are relatively stagnant and remain either on or behind the stenosis for the entire cycle. Representative trajectories of these three scenarios are presented in Fig. 4, where the fluid particle's trajectory and its initial positions are represented by full black lines and a black full circle, respectively. The green star represents the particle's position for each snapshot. Trajectories \textit{A} and \textit{B} correspond to fluid particles that were pushed downstream by the incoming flow. Trajectory \textit{C} represents a fluid particle that was pushed downstream and then dragged by the vortex. Finally, trajectories \textit{D} and \textit{E} represent stagnant fluid particles on and behind the stenosis, respectively. 

\begin{figure*}[t!]
\includegraphics[width = 1 \linewidth]{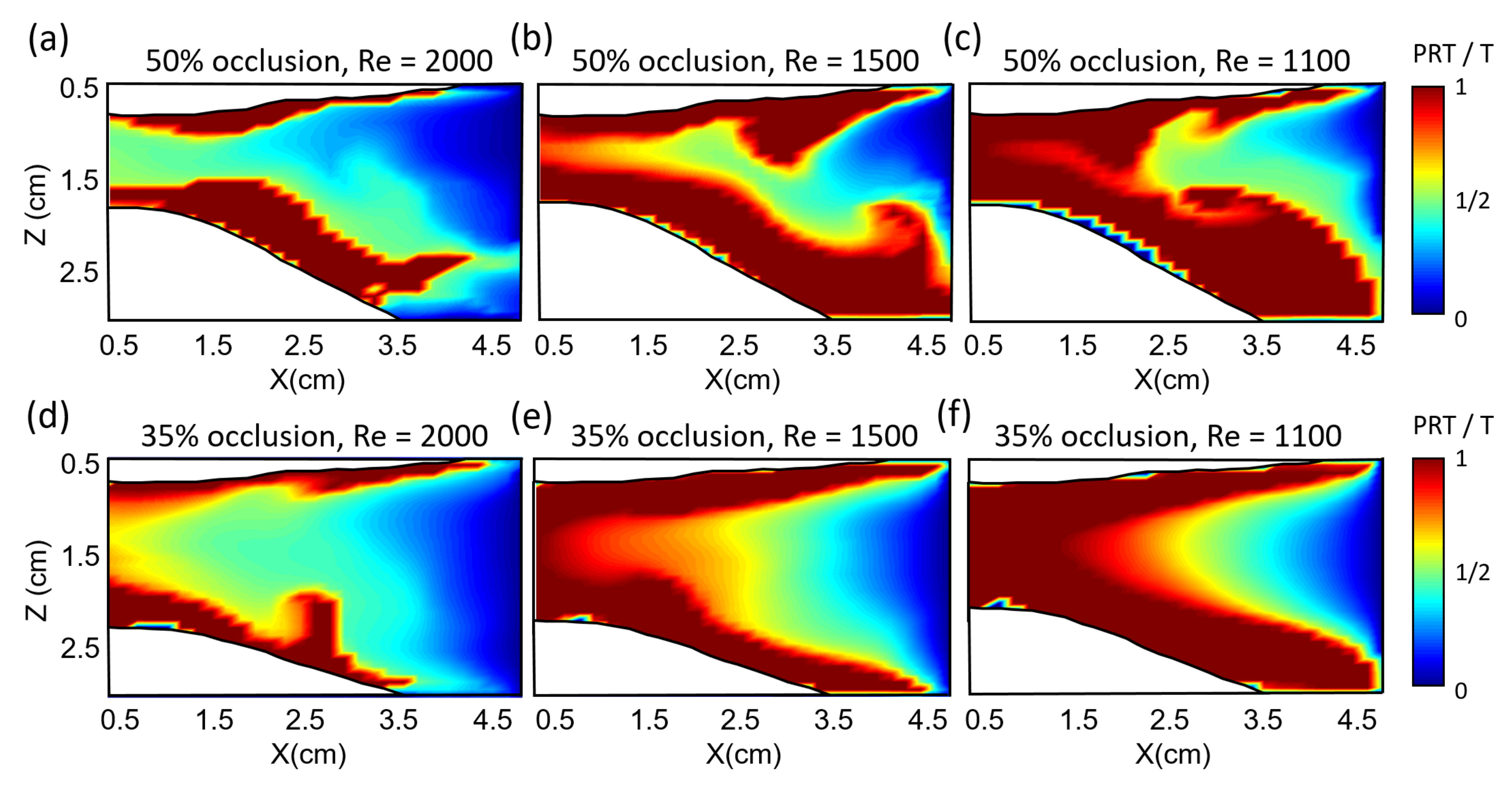}
\caption{\label{fig:PRT} Particle Residence Time (PRT) maps for the following experiments \textbf{(a)} 50\% occlusion, $Re$ = 2000, \textbf{(b)} 50\% occlusion, $Re$ = 1500, \textbf{(c)} 50\% occlusion, $Re$ = 1100, \textbf{(d)} 35\% occlusion, $Re$ = 2000, \textbf{(e)} 35\% occlusion, $Re$ = 1500 and \textbf{(f)} 35\% occlusion, $Re$ = 1100. Each pixel corresponds to the fluid particle's initial position and the color value to the time spent by the particle within the ROI normalized by the pulsatile period $T$.}
\end{figure*}

\subsubsection{Residence time and final position maps}
\label{prt_results}

To evaluate these different flow behaviours we computed the PRT (Fig. 5) and the FP maps (Fig. 6). For the unobstructed model (not shown) no stagnant regions were observed. All particles travelled across the ROI following an approximately straight path oriented along x-axis. The time for a particle to travel across the ROI was $\approx T$ while the average PRT over the ROI was $\approx T/2$. 

For the occluded models flow behaviour was quite different. Figure 5 shows the PRT maps for the different $Re$ in the occluded models.  From Fig. 5 it is possible to identify the two main flow structures: the high velocity jet with low PRT values ($< T/2$ ) and the recirculation region (on and behind the stenosis) with high PRT values ($> T/2$). For a given $Re$, the PRT of the fluid particles initially situated in the recirculation region increases with the degree of occlusion. For example, for $Re = 2000$, we observe an overall increase in the PRT by comparing the 50\% (Fig. 5(a)) to the 35\% (Fig. 5(d)) occluded model. Moreover, we observe that the area occupied by high PRT values (i.e. close to $T$) is larger for the 50\% occluded model. This is consistently observed for the rest of the $Re$ by comparing Figs. 5(b) to 5(e) and Figs. 5(c) to 5(f). Lastly, for a given degree of occlusion (Figs. 5(a) to 5(c) for a 50\% occlusion and Figs. 5(d) to 5(f) for a 35\% occlusion), the  PRT over the ROI increases as $Re$ decreases.

Since Fig. 5 only gives us information about the time a fluid particle spent within the ROI without any information on its position, we complete this temporal information by introducing the FP maps (Fig. 6). For the FP maps the ROI was subdivided into three different subregions representing: (i) the region occupied by the jet, (ii) the region on the stenosis and (iii) the region behind the stenosis. Those regions are indicated by white dashed lines in Fig. 6. More precisely, region $i$ goes from the anterior wall of the model (i.e. proximal to the ultrasonic array) to the stenosis throat, occupying the whole post-stenotic region along the x-direction. Regions $ii$ and $iii$ both go from the stenosis throat to the posterior wall. However, region $ii$ goes from X  $\approx$ 0.4 cm to X $\approx$ 3 cm, while region $iii$ goes from X $\approx$ 3 cm to X $\approx$ 4.8 cm, representing the regions on and behind the stenosis, respectively. The choice of these regions allowed us to evaluate the three different scenarios described above for Fig. 4: fluid particles pushed by the flow, fluid particles dragged by the vortex and stagnant particles that stayed on and behind the stenosis. In Fig. 6 the colours cyan, orange and red correspond to a fluid particle's final position within regions $i$, $ii$ and $iii$, respectively, while the color blue is for particles that have left the region.  

In Fig. 6, we observe that part of the fluid particles with high PRT values, which initially were located on the stenosis (region $ii$), moved to region $iii$ while the rest remained on the stenosis (i.e. region $ii$). The number of particles remaining on the stenosis does not strongly depend on $Re$. By comparing Figs. 6(a) to 6(c) and Figs. 6(d) to 6(f) we observe that the size of the orange surface remains approximately the same. However, we observe that the lower the $Re$ the more particles remain in region $i$ and $iii$. Finally, we also observe some cyan regions within region $ii$ for Figs. 6(a) and 6(b). These regions correspond to fluid particles that were dragged into region $i$ by the vortex.

\begin{figure*}[!]
\includegraphics[width = 1 \linewidth]{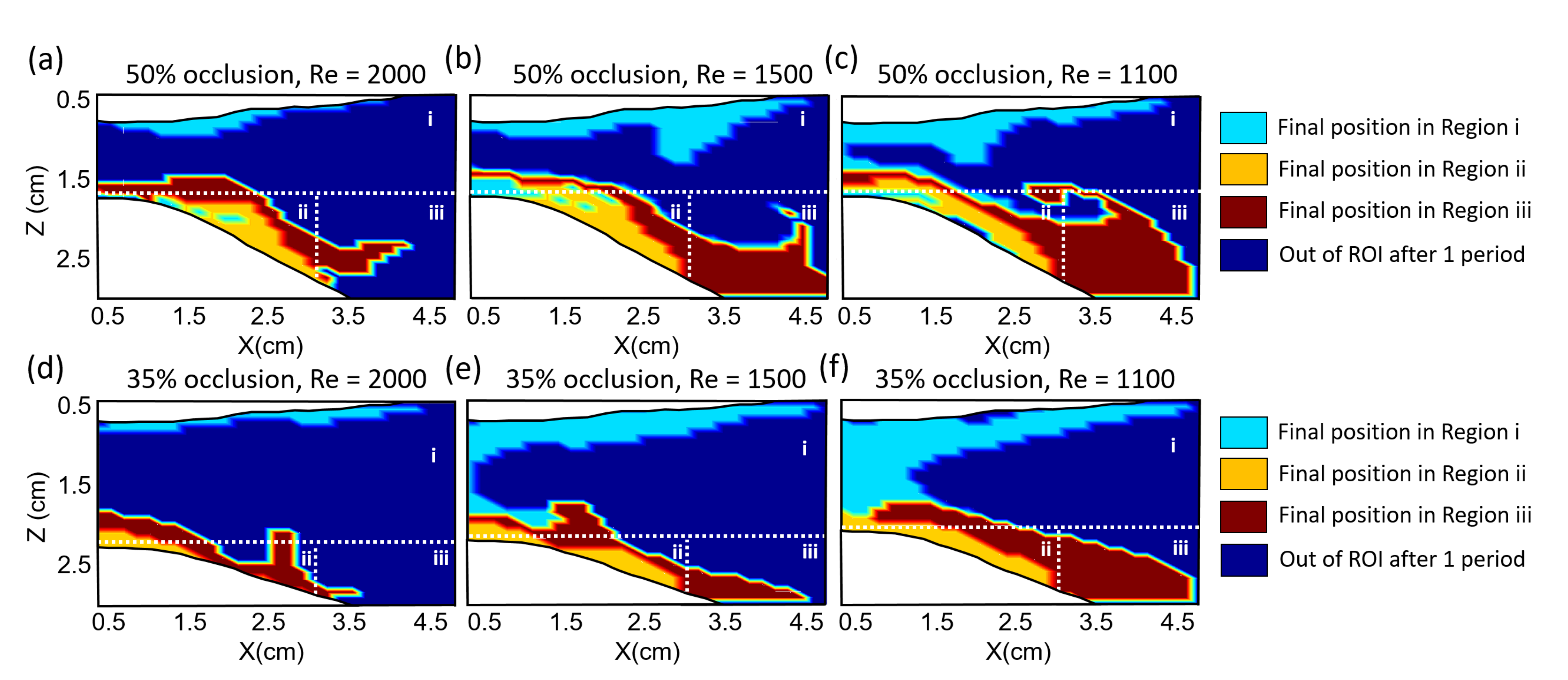}
\caption{\label{fig:FP} Final Position (FP) maps for the following experiments: \textbf{(a)} 50\% occlusion, $Re$ = 2000, \textbf{(b)} 50\% occlusion, $Re$ = 1500, \textbf{(c)} 50\% occlusion, $Re$ = 1100, \textbf{(d)} 35\% occlusion, $Re$ = 2000, \textbf{(e)} 35\% occlusion, $Re$ = 1500 and \textbf{(f)} 35\% occlusion, $Re$ = 1100. Each pixel corresponds to the fluid particle's initial position and the color value corresponds to the particle's final position within a given subregion: colours cyan, orange and red correspond to subregion $i$, $ii$ and $iii$, respectively, which are indicated by a dashed white line. The color blue is for particles that left the region in one period.}
\end{figure*}

\subsubsection{FTLE fields}

Figure 7 shows six snapshots of the instantaneous ridges of the FTLE fields superimposed to the instantaneous velocity field for different experiments. Each snapshot corresponds to the time reference set in Fig. 2(d). Blue and red ridges corresponds to the attracting ($\Lambda^+$) and repelling ($\Lambda^-$) LCSs respectively. We observed several LCSs related to the high velocity jet and the vortex. The LCS indicated by $I$ is related to the leading edge of the central jet. This attracting LCS separates the incoming flow from the pre-existing fluid. We observe that the intensity and the extension of this barrier increases with $Re$. Moreover, the LCSs denoted by $IIa$ and $IIb$ in Fig. 7 are related to the vortex. Both LCSs surround the vortex at all times and are the main responsible for the mixing behind the stenosis. Fluid particles trapped between those barriers travel along the vortex, repelled by LCS $IIa$ (red) and simultaneously attracted by LCS $IIb$ (blue). Moreover, we observe that the strength of both barriers increase with $Re$ while the overall structure is maintained. Finally, an additional vortex on the anterior wall of the model is observable in Figs. 7(a) and 7(b). This vortex is more evident from the FTLE fields than from the vorticity field presented in Figs 3(a) and 3(b).  

\begin{figure*}[t!]
\includegraphics[width = 1 \linewidth]{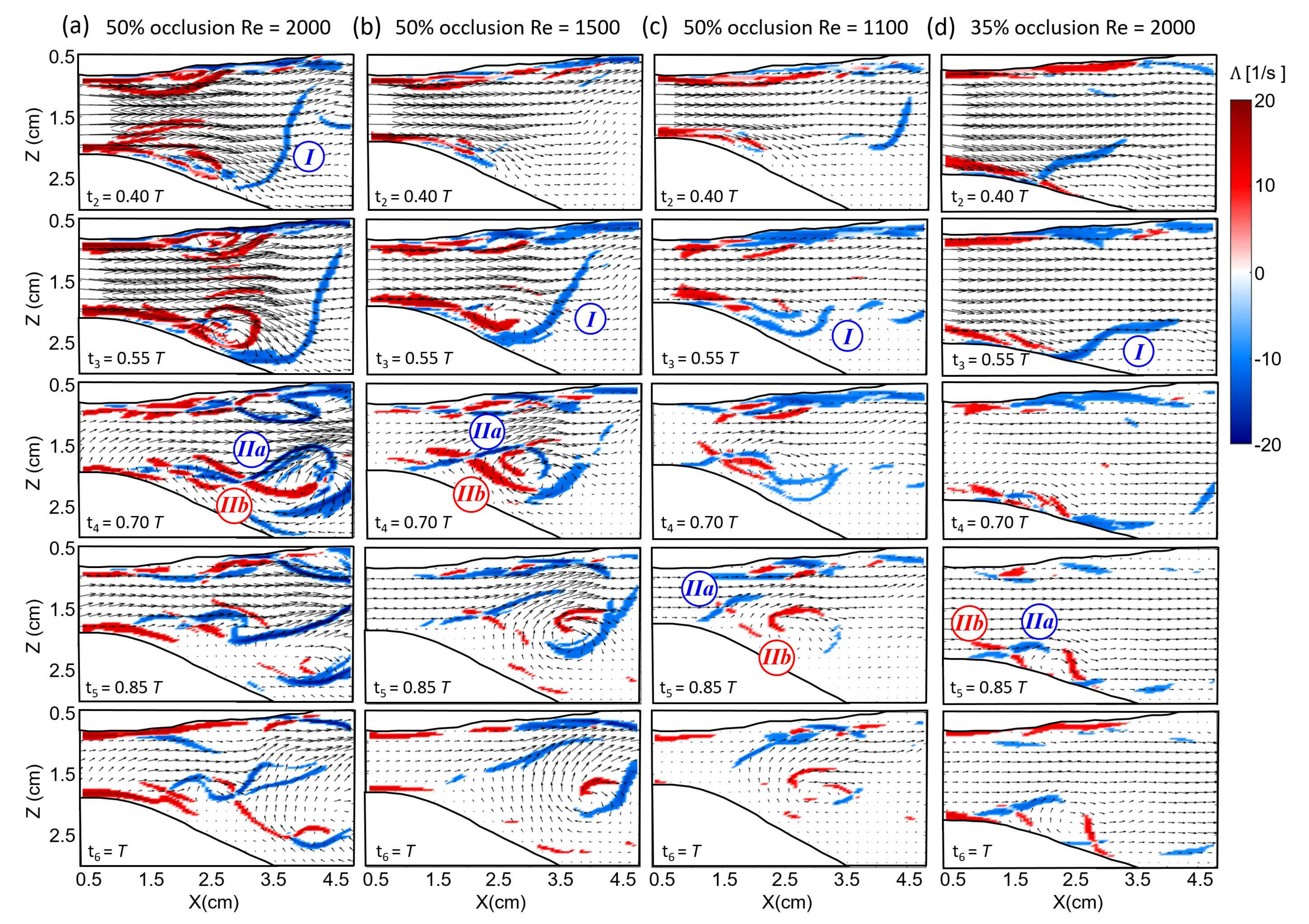}
\caption{\label{fig:FTLE} Six snapshots of the instantaneous ridges of the FTLE fields superimposed to the instantaneous particle velocity field for the following experiments \textbf{(a)} 50\% occlusion, Re = 2000, \textbf{(b)} 50\% occlusion, Re = 1500, \textbf{(c)} 50\% occlusion, Re = 1100 and \textbf{(d)} 35\% occlusion, Re = 2000. Each snapshot corresponds to the times $t_2$ to $t_6$ indicated in Fig. 2(d). Attracting ($\Lambda^+$) and repelling ($\Lambda^-$) LCSs are represented in blue and red respectively. The color-scale is the same for all panels.}
\end{figure*} 

\section{Discussion}

In this work a novel approach which combines ultrasound with Eulerian and Lagrangian descriptors has been applied to study vortex dynamics and transport phenomena in stenotic aortic models with morphology, mechanical and optical properties close to those of real arteries. We observed that the flow in the post-stenotic region is composed of two main structures: a high velocity jet over the stenosis throat and a recirculation region behind the stenosis where vortex form and shed. This result is consistent with previous numerical and experimental studies conducted in axis-symmetric \cite{varghese2007direct,geoghegan2013time,pielhop2012analysis} and non-symmetrical models \cite{usmani2016pulsatile,choi2018flow}. 

For those experiments where vortex shedding was observed, we were able to track the vortex center from the vorticity snapshots and  measure its axial velocity of propagation. Although we could not find any values reported in the literature for the vortex velocity of propagation in stenotic vessels, we were able to deduce them from the work of Geoghegan \textit{et al.} \cite{geoghegan2013time}. The values obtained in our work are smaller by factor $\sim$10 when compared to the results deduced from \cite{geoghegan2013time}. However, those experiments were conducted in an axis-symmetrically occluded silicone model. Since the $Re$ used in our experiments and in \cite{geoghegan2013time} are similar, the difference between the results is probably related to the elasticity difference between models, since silicone is far less compliant than PVA. Moreover, in their work they report a Kelvin-Helmholtz instability, which we do not observe in our experiments due to the asymmetry of our model.  This is consistent with \cite{choi2018flow,usmani2016pulsatile}.

By computing the FTLE fields (Fig. 7), we were able to identify different material barriers associated to the vortex and the leading edge of the jet. Analogous LCSs were reported by previous studies but in significantly different hemodynamic situations. For example, in \cite{hendabadi2013topology} and \cite{espa2012lagrangian} FTLE fields were computed to study transport in the left ventricle. In those studies, they found that during the left ventricle filing the backward FTLE field reveals a well-defined LCS that separates the injected blood from the pre-existing one. Other numerical studies in abdominal aortic aneurysm showed analogous behavior during diastole \cite{shadden2008LCS_Vessels,arzani2012FTLE_AAA}. The LCS described in these works is analogous to the attracting barrier $I$ identified in Fig. 7. However, contrary to what is described in   \cite{espa2012lagrangian,hendabadi2013topology,shadden2008LCS_Vessels,arzani2012FTLE_AAA}, due to the asymmetric nature of our models and flow, the LCS $I$ does not roll originating a vortex ring whose trailing edge (i.e. facing upstream) is enclosed by a repelling barrier. This is evident from Fig. 7(a) for times $t_5$ and $t_6$: as vortex leave the post-stenotic region there are still ridges of the FTLE field separating the jet from the recirculation region.

In this work we also identified material barriers associated to the vortex (LCSs $IIa$ and $IIb$ in Fig. 7). Both LCSs surround the vortex at all times and are the main responsible for the mixing between the incoming and the pre-existing fluid as described in \cite{shadden2007transport}. By comparing Figs. 4 and 7(b) we observed that during vortex build up, pre-existing fluid gets trapped between barriers $IIa$ and $IIb$ and travel along the vortex getting mixed and stirred with the incoming flow. Similar dynamics were also observed for $Re =$ 2000 and 1100, although it was not illustrated in Fig. 4. For the 35\% occluded model at $Re = 2000$ those barriers were also visible around the vortex (specially for times $t_5$ and $t_6$ in Fig. 7(d)). Analogous results were reported by computing the FTLE fields from numerical studies in aortic aneurysm \cite{arzani2012FTLE_AAA,shadden2008LCS_Vessels} and in the experimental work of Vetel \textit{et al.} in a rigid model of a carotid bifurcation \cite{vetel2009lagrangian}. Lastly, in this work we showed that the strenght of this barriers increases with $Re$ and consequently with vorticity. 

To complete the information conveyed by the FTLE fields we computed the PRT and the FP maps. Particularly, PRT and FP maps as defined in this work, provide detailed information on the transport of the pre-existing fluid (blue dots in Fig. 4). By comparing Figs. 5 and 6, we observed that part of the fluid  initially located on the stenosis (region $ii$), moved to region $iii$ while the rest remained stagnant on the stenosis. In Fig. 6 we observed that the number of these stagnant particles does not strongly depend on $Re$ for the range of values explored in this work. Contrary, results reported by Jerónimo \textit{et al.} indicate that the number of stagnant particles increases as $Re$ increases \cite{jeronimo2020stenosis}, however, comparison between their results and the ones reported in this work is not straightforward. In \cite{jeronimo2020stenosis} they used peak $Re$ ranging from 7200 to 28080 in a rigid smooth axisymmetric constriction followed by an unrealistic sudden expansion, which is quite different from the setup used in this work. 

As in \cite{jeronimo2020stenosis}, our experiments were also limited by the acquisition time of the ultrasound scanner and the velocity imaging through Echo-PIV was restricted to the centre plane of the model. The current setup did not allow the study of the three dimensional nature of the flow. Future studies should aimed in incorporating three dimensional information on vortical structures and transport phenomena. This can be implemented with the advent of new 3D ultrasound imaging technologies \cite{jensen2016ultrasound_II} like matrix \cite{correia20164d} and row-columns \cite{sauvage2018large} arrays. 

Lastly, LCS, PRT and FP maps may provide an alternative approach to understand stenosis growth. As shown by \cite{shadden2013potential} platelet activation potential is maximized along the LCSs. Moreover, from the PRT and FP maps we showed that stagnant particles remained on stenosis after one pulsatile period. These stagnant particles spend more time on the stenosis where platelet activation potential is large (i.e. large FTLE values as in Fig. 7) which may favour blood clot formation leading to stenosis growth.

\section{Conclusion}

By using a novel approach which combines ultrasound with Eulerian and Lagrangian descriptors, we analysed the flow dynamics and transport phenomena in stenotic aortic models with morphology, mechanical and optical properties close to those of real arteries. To this end, vorticity, FTLEs, PRT and FP maps were computed from the particle velocity fields acquired using Echo-PIV. We characterized vortex kinematics showing that vortex propagation velocity increases with $Re$ number. From the FTLE field we identified material barriers that dictate transport behind the stenosis. The size and strength of these barrier is also $Re$ depend. Lastly, from the PRT and FP maps, we showed that even for the highest $Re$, fluid parcels remain on the stenosis which combined with large FTLE values may provide an alternative way to understand the process of stenosis growth. This comprehensive study proves that this type of approach may help to bridge the gap between fundamental physics and relevant clinical applications such as plaque evolution and rupture.

\ack

This work was supported by CSIC I+D 2016 Project ``Estudio de la dinamica de un flujo pulsatil y sus implicancias en hemodinamica vascular'' Uruguay, ANII-Uruguay, PEDECIBA Física - Uruguay, the Administrative Department of Science, Technology and Innovation of the Colombian Government (Colciencias) and the Research and Development Center (CIDI) of the Universidad Pontificia Bolivarina through the program ``Es tiempo de Volver'' (Grant number 548B-01/16-04), Colombia. Nicasio Barrere acknowledges a doctoral scholarship (POSNAC-2015-1-109843) from ANII-Uruguay.


\section*{References}


\bibliography{FTLE_aortic_models}

\end{document}